\documentclass[aps,twocolumn,pra,showpacs,amsmath,amssymb,showkeys,10pt]{revtex4-2}
\usepackage{graphicx}
\usepackage{epstopdf}
\usepackage{diagbox}
\pdfoutput=1
\usepackage{braket}
\usepackage{hyperref}
\usepackage{longtable}

\begin{document}

	\title{Distributed computing quantum unitary evolution}
	
	\author{Hui-hui Miao}
	\email[Email address: ]{hhmiao@cs.msu.ru}
	\affiliation{Faculty of Computational Mathematics and Cybernetics, Lomonosov Moscow State University, Vorobyovy Gory 1, Moscow, 119991, Russia}

	\author{Yuri Igorevich Ozhigov}
	\email[Email address: ]{ozhigov@cs.msu.ru}
	\affiliation{Faculty of Computational Mathematics and Cybernetics, Lomonosov Moscow State University, Vorobyovy Gory 1, Moscow, 119991, Russia\\K. A. Valiev Institute of physics and technology, Russian Academy of Sciences, Nakhimovsky Prospekt 36, Moscow, 117218, Russia}

	\date{\today}

	\begin{abstract}
	A distributed computing approach to solve the curse of dimensionality, caused by the complex quantum system modeling, is discussed. With the help of Cannon's algorithm, the distributed computing transformation of numerical method for simulating quantum unitary evolution is achieved. Based on the Tavis--Cummings model, a large number of atoms are added into the optical cavity to obtain a high-dimensional quantum closed system, implemented on the supercomputer platform. The comparison of time cost and speedup of different distributed computing strategies is discussed.
	\end{abstract}

	%\pacs{03.65.Yz, 42.50.Lc, 42.50.-p, 42.50.Pq}

	\maketitle

	\section{Introduction}
	\label{sec:Intro}
	
	Complex quantum system modeling is one of the most important directions in computational mathematics today, especially in the computational fields involving polymer chemistry and macromolecular biology \cite{McArdle2020, Baiardi2023, Albuquerque2021}. When simulating chemical and biological reactions, a large number of particles are often involved, and the dimension of the quantum system composed of these particles increase exponentially as the number of particles increases, thus causing a problem called the curse of dimensionality --- raised by R.E. Bellman \cite{Bellman1957, Bellman1961} to describe a number of events that occur while organizing and evaluating data in high-dimensional areas. The curse of dimensionality has always been a major obstacle to the study of high-dimensional quantum systems. However, the study of the structure of biological macromolecules is one of the most important frontier studies of quantum computing, thus the curse of dimensionality is an urgent issue to be solved. With the development of supercomputers in recent decades, the use of distributed computing can solve a series of memory and efficiency problems caused by the curse of dimensionality to a certain extent. Moreover, some distributed computing algorithms can be applied to quantum unitary evolution.
	
	A key contribution of this paper is the cavity quantum electrodynamics (QED) model, which is easy to implement in the laboratory and offers a unique scientific paradigm for studying light-matter interaction. According to this paradigm, impurity two-level systems, also known as atoms, are connected to fields of cavities. The cavity QED model includes the Jaynes--Cummings model (JCM) \cite{Jaynes1963} and the Tavis-Cummings model (TCM) \cite{Tavis1968} as well as their generalizations. Many studies have been conducted recently in the field of these models \cite{Prasad2018, Guo2019, Kulagin2022, Miao2023, Ozhigov2023}. In our previous studies, some methods have been used to solve some of the computing problems caused by the curse of dimensionality \cite{You2023, Chen2023}.
	
	In this paper, we try to simulate the cavity QED model with a large number of two-level artificial atoms on a supercomputing platform to reveal the rules of unitary evolution of a complex quantum system, and study the time cost and speedup under different distributed computing strategies comparing with that under the situation without distributed computing. This paper is organized as follows. After introducing the TCM with a large number of artificial atoms in Sec. \ref{sec:TCModel}, we introduce numerical method and its distributed computing transformation in Sec. \ref{sec:Method}. We present the results of our numerical simulations in Sec. \ref{subsec:Interaction} and study the efficiency of distributed computing method in Sec. \ref{subsec:Comparison}. Some brief comments on our results in Sec. \ref{sec:Conclusion} close out the paper.
	
	\section{Tavis--Cummings model with a large number of artificial atoms}
	\label{sec:TCModel}
	
	\begin{figure*}
		\centering
		\includegraphics[width=0.7\textwidth]{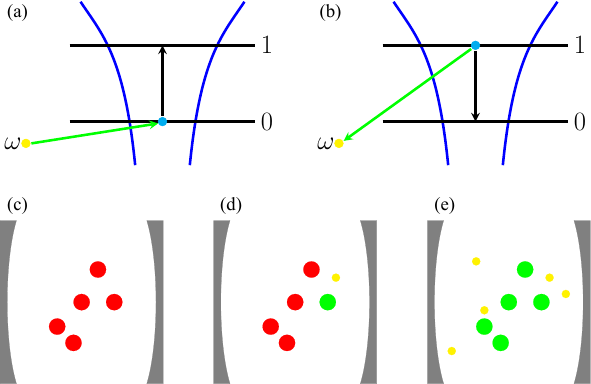} 
		\caption{(online color) Schematic diagram of Tavis--Cummings model with a large number of artificial atoms. (a) and (b) represent in detail the excitation and de-excitation processes of a two-level artificial atom, respectively. (c) $\to$ (d) $\to$ (e) represents the de-excitation of atoms, (e) $\to$ (d) $\to$ (c) represents the excitation of atoms. Electron, photon, excited atom and ground state atom are as blue, yellow, red and green dot, respectively.}
		\label{fig:TCModel}
	\end{figure*}
	
	In this paper, we introduce the TCM involving a large number of atoms. The basic states of the TCM is as follow
	\begin{equation}
		\label{eq:BasisTCModel}
		|\Psi\rangle=\bigotimes_{i=1}^{n}|p_i\rangle|l_i\rangle
	\end{equation}
	where $n$ --- number of atoms, $p_i\in[0,1]$ --- number of free photons from i-st atom, $l_i\in[0,1]$ --- electronic state, $l_i=0$ --- electron in ground state in i-st atom, $l_i=1$ --- electron in excited state in i-st atom. Since the closed system, we can omit the qubit involving photons
	\begin{equation}
		\label{eq:BasisTCModelSimple}
		|\Psi\rangle=\bigotimes_{i=1}^{n}|l_i\rangle
	\end{equation}
	And we can intuitively get that  the dimension of Hilbert space is $N=2^n$.
	
	Interaction between atom and field is explained in detail in Fig. \ref{fig:TCModel}. In (a), an electron in the ground state absorbs a photon and transfer to the excited state, which is called excitation. In (b), an electron in the excited state transfer to the ground state after releasing a photon, this process is called de-excitation. Initial state is shown in Fig. \ref{fig:TCModel} (c), where exist $n$ excited atoms and no photons. The excited atom will de-excite and become a ground state atom, at which time a photon will be released (see Fig. \ref{fig:TCModel} (d)). When all atoms change to the ground state, $n$ photons are present in the optical cavity (see Fig. \ref{fig:TCModel} (e)). In a closed system, the de-excitation and excitation of atom exist at the same time, so a certain regularity will appear in the light-matter system, which is embodied in the periodic oscillation of the quantum states.
	
	Before constructing the Hamiltonian, we first introduce rotating wave approximation (RWA) \cite{Wu2007}, which is taken into account
	\begin{equation}
		\label{eq:RWACondition}
		\frac{g}{\hbar\omega_a}\approx\frac{g}{\hbar\omega_c}\ll 1
	\end{equation}
	 Usually, for convenience, we assume that the electron transition frequency $\omega_a$ and field frequency $\omega_c$ are equal, and $\omega=\omega_a=\omega_c$. Now Hamiltonian of TCM in the case of RWA has following form
	\begin{equation}
		\label{eq:Hamiltonian}
		H_{TCM}^{RWA}=\hbar\omega a^{\dag}a+\sum_{i=1}^n\left[\hbar\omega\sigma_i^{\dag}\sigma_i+g_i\left(a^{\dag}\sigma_i+a\sigma_i^{\dag}\right)\right]
	\end{equation}
	where $\hbar$ is the reduced Planck constant, $g_i$ is the coupling intensity between the field and the electron in the atom. Here $a$ is photon annihilation operator, $a^{\dag}$ is photon creation operator, $\sigma_i$ is electron relaxation operator, and $\sigma_i^{\dag}$ is electron excitation operator.
	
	\section{Numerical method and its distributed computing transformation}
	\label{sec:Method}
	
	\subsection{Numerical method}
	\label{subsec:Method}
	
	\begin{figure*}
		\centering
		\includegraphics[width=1.\textwidth]{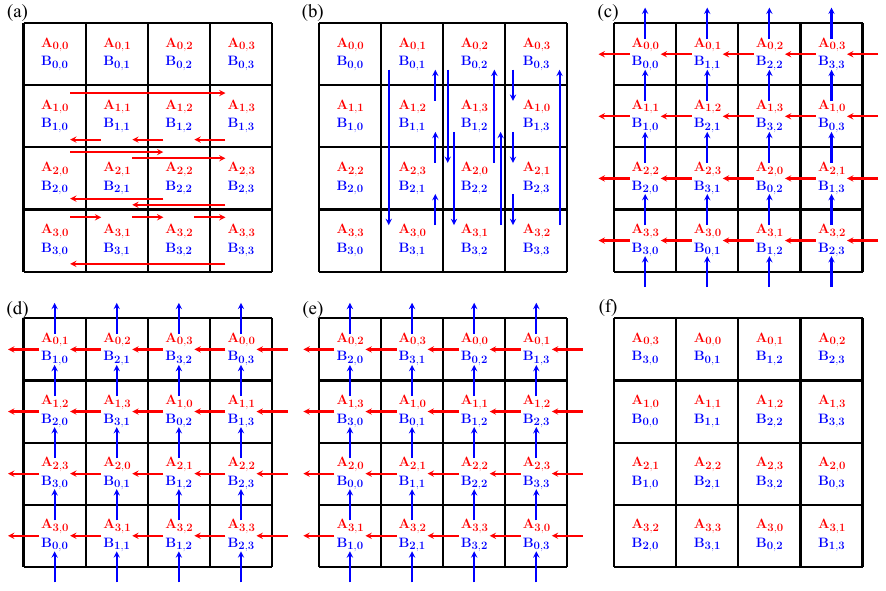} 
		\caption{(online color) Schematic diagram of Cannon's algorithm. (a) and (b) are alignmentation operations, (c), (d) and (e) are transferring operations.}
		\label{fig:Cannon}
	\end{figure*}
	
	The quantum master equation (QME) in the Markovian approximation for the density operator $\rho$ of the system takes the following form
	\begin{equation}
		\label{eq:QME}
		i\hbar\dot{\rho}=\left[H,\rho\right]+iL\left(\rho\right)
	\end{equation}
	where $H$ --- Hamiltonian. And the Lindblad term $L\left(\rho\right)$ involving dissipation process, which is temporarily disabled in this paper, can be deleted. Thus, the solution $\rho\left(t\right)$ in Eq. \eqref{eq:QME} can be approximately found as follows
	\begin{equation}
		\label{eq:UnitaryPart}
		\rho\left(t+dt\right)=exp\left(-\frac{i}{\hbar}Hdt\right)\rho\left(t\right)exp\left(\frac{i}{\hbar}Hdt\right)
	\end{equation}
	
	In order to apply distributed computing algorithm, the matrix exponential parts in the Eq. \eqref{eq:UnitaryPart} needs to be converted into matrix multiplication-addition operations. In the past few decades, many calculation methods for exponential matrices have been proposed \cite{Moler2003, Sidje1998}. Here we use the Taylor series approximation
	\begin{equation}
		\label{eq:Taylor}
		exp(A)=\sum_{k=0}^{\infty}\frac{A^k}{k!}
	\end{equation}
	where $A$ is matrix. Now Eq. \eqref{eq:UnitaryPart} can be rewritten as follows
	\begin{equation}
		\label{eq:UnitaryPartTaylor}
		\rho\left(t+dt\right)=\left[\sum_{k=0}^{\infty}\frac{(-\frac{i}{\hbar}Hdt)^k}{k!}\right]\rho\left(t\right)\left[\sum_{k=0}^{\infty}\frac{(\frac{i}{\hbar}Hdt)^k}{k!}\right]
	\end{equation}
	where $k_{max}=10$, which is enough to ensure accuracy.
	
	\subsection{Cannon's algorithm}
	\label{subsec:Cannon}
	
	In this paper, we use Cannon's algorithm \cite{Cannon1969} to solve the memory problem caused by large matrices with high dimension. This algorithm is used to process multiplication-addition operations of large matrices. Its core idea is to divide the matrix into many small blocks and send them to processors. Each processor receives its own block in sequence. This method can significantly reduce the occupied memory of each processor, but it also causes some disadvantages: as the number of processors increases, the time cost required for data transmission across processors and across nodes also increases.
	
	According to Eq. \eqref{eq:BasisTCModelSimple}, the dimension of Hamiltonian and density matrix is equal to $N=2^n$. We suppose the number of processors is $p$, and $p_x=p_y=\sqrt{p}$. Thus, a matrix can be divided into $p_x\times p_y$ blocks. At the beginning, block on the processor $P_{i,j}$ is denoted by $A_{i,j}$ (or $B_{i,j}$), where $0\le i,j<p_x$. The steps of Cannon's algorithm are as follows
	\begin{itemize}
		\item Generalization: generate block $A_{i,j}$ from matrix $A$ and block $B_{i,j}$ from matrix $B$ on processor $P_{i,j}$;
		\item Horizontal alignmentation: shift block $A_{i,j}$ to the left $i$ steps (see Fig. \ref{fig:Cannon} (a));
		\item Vertical alignmentation: shift block $B_{i,j}$ up $j$ steps (see Fig. \ref{fig:Cannon} (b));
		\item Transferring: firstly, perform multiplication-addition operations on each processor, then shift its block from matrix $A$ one step to the left and get the another block from his right neighbour cyclically, shift block from matrix $B$ up one step and get the another block from the neighbour below him cyclically;
		\item Repeat transferring operation $p_x$ times and perform multiplication-addition operations on each processor, finally get the product of matrices $A$ and $B$ (see Fig. \ref{fig:Cannon} (c) $\sim$ (f)).
	\end{itemize}
	
	We suppose function $Cannon(M_1,M_2)$ is the code of Cannon's algorithm for two arbitrary square matrices $M_1$, $M_2$ with the same dimension. Similarly, we have
	\begin{widetext}
		\begin{subequations}
			\label{eq:CannonFunc}
			\begin{align}
				&Cannon^2(M_1,M_2)=Cannon(M_1,M_2)\label{eq:CannonFunc2}\\
				&Cannon^3(M_1,M_2,M_3)=Cannon(M_1,Cannon(M_2,M_3))\label{eq:CannonFunc3}\\
				&Cannon^4(M_1,M_2,M_3,M_4)=Cannon(M_1,Cannon(M_2,Cannon(M_3,M_4)))\label{eq:CannonFunc4}\\
				&\cdots\cdots\nonumber\\
				&Cannon^k(M_1,M_2,\cdots,M_{k-1},M_k)=Cannon(M_1,Cannon(M_2,\cdots,Cannon(M_{k-1},M_k)))\label{eq:CannonFunck}
			\end{align}
		\end{subequations}
	\end{widetext}
	If $M_1=M_2=\cdots=M_{k-1}=M_k=M$, $Cannon^k(M_1,M_2,\cdots,M_{k-1},M_k)$ can be simply rewritten as $Cannon^k(M)$.
	
	After construction of Cannon's algorithm, we can transform the Eq. \eqref{eq:UnitaryPartTaylor} to a distributed computing form. In the first step, we get left matrix exponential part in Eq. \eqref{eq:UnitaryPartTaylor} through Taylor series approximation
	\begin{equation}
		\label{eq:TaylorLeft}
		L=I-\frac{i}{\hbar}Hdt+\sum_{k=2}^{10}\frac{Cannon^k(-\frac{i}{\hbar}Hdt)}{k!}
	\end{equation}
	where $I$ is unit matrix, and we suppose maximum of exponents is $10$ in this paper. Then in the same way, we get right matrix exponential part in Eq. \eqref{eq:UnitaryPartTaylor} through Taylor series approximation
	\begin{equation}
		\label{eq:TaylorRight}
		R=I+\frac{i}{\hbar}Hdt+\sum_{k=2}^{10}\frac{Cannon^k(\frac{i}{\hbar}Hdt)}{k!}
	\end{equation}
	Finally, we transform Eq. \eqref{eq:UnitaryPartTaylor} into the following form
	\begin{equation}
		\label{eq:UnitaryPartTaylorCannon}
		\rho\left(t+dt\right)=Cannon^3(L,\rho\left(t\right),R)
	\end{equation}
	Now distributed computing transformation of numerical method of quantum unitary evolution is completed.
	
	\section{Results}
	\label{sec:Results}
	
	\begin{figure}
		\centering
		\includegraphics[width=0.5\textwidth]{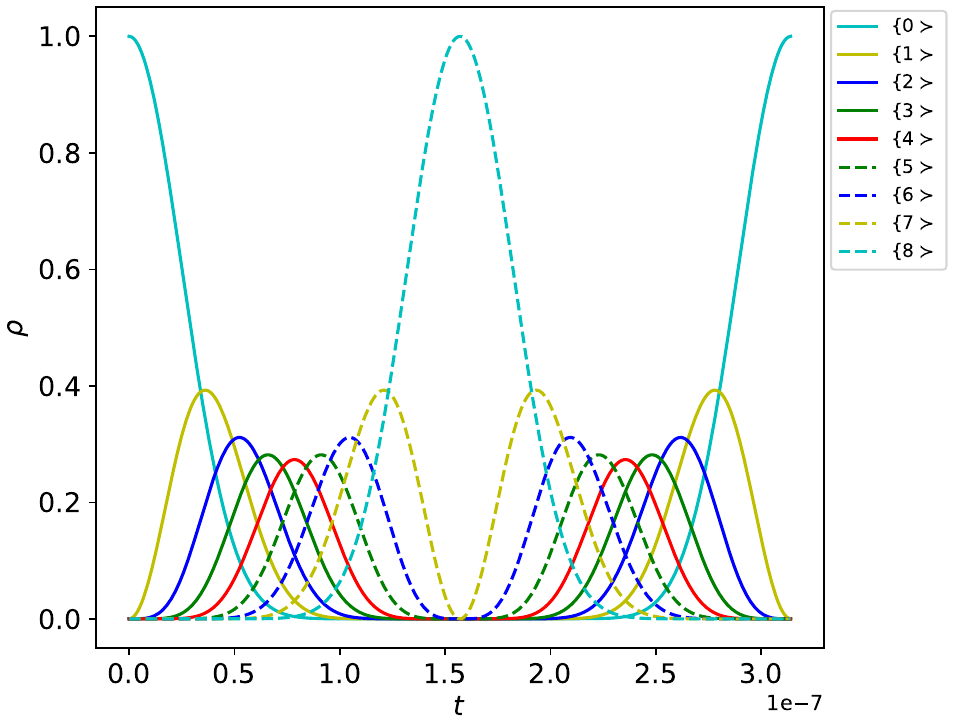} 
		\caption{(online color) Unitary evolution of TCM with 8 atoms.}
		\label{fig:Evolution8}
	\end{figure}
	
	\subsection{The interaction between photons and matter}
	\label{subsec:Interaction}
	
	\begin{figure*}
		\centering
		\includegraphics[width=1.\textwidth]{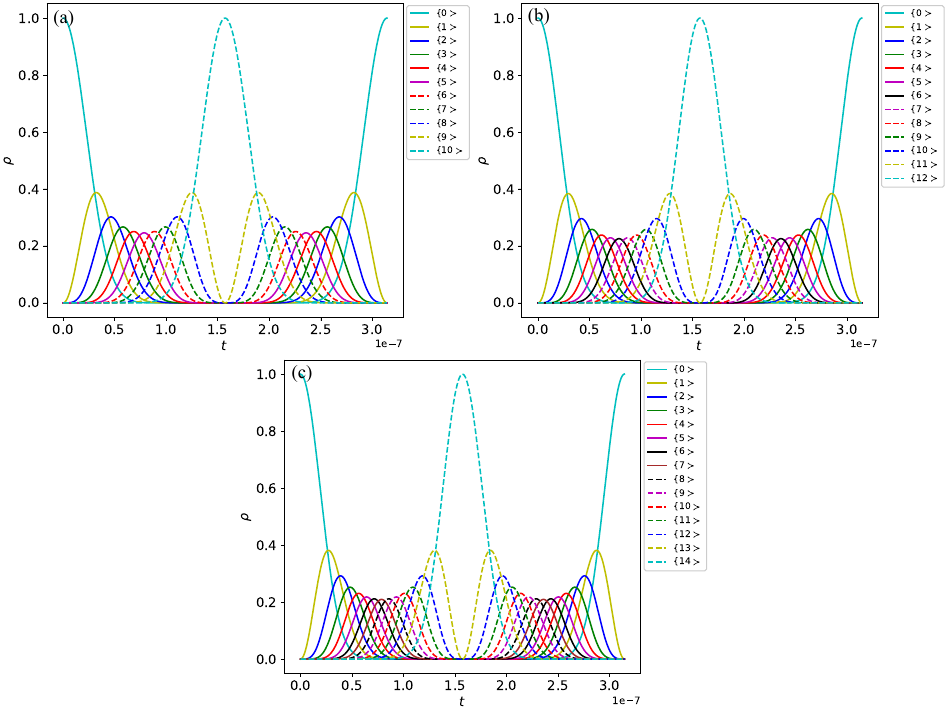} 
		\caption{(online color) Unitary evolution of TCM with more atoms. (a), (b) and (c) represent unitary evolution of TCM with 10, 12 or 14 atoms, respectively.}
		\label{fig:EvolutionMore}
	\end{figure*}
	
	\begin{table*}[!htpb]
        \centering
		\begin{tabular}{|c|c|c|c|c|c|c|c|c|}
			\hline
			\diagbox{Strategy}{Time}{Dimension} & $2^8$ & $2^9$ & $2^{10}$ & $2^{11}$ & $2^{12}$ & $2^{13}$ & $2^{14}$ & $2^{15}$ \\
			\hline
			without MPI & $<$1s & $<$1s & 3s & 22s & 2m34s & -- & -- & -- \\
			\hline
			$2\times2$ & $<$1s & $<$1s & 1s & 8s & 1m6s & 8m4s & -- & -- \\
			\hline
			$4\times4$ & $<$1s & 1s & 5s & 20s & 1m26s & 6m45s & 34m54s & -- \\
			\hline
			$8\times8$ & 2s & 6s & 20s & 44s & 2m35s & 9m43s & 38m14s & 3h58m20s \\
			\hline
			$16\times16$ & 6s & 20s & 29s & 1m34s & 3m26s & 24m55s & 42m46s & 3h16m2s \\
			\hline
		\end{tabular}
		\caption{Comparison of time cost of distributed computing Taylor series approximation. The symbols "h", "m" and "s" represent "hour(s)", "minute(s)" and "second(s)", respectively.}	
		\label{Tab:ComparisonTimeTaylor}
	\end{table*}
	
	\begin{table*}[!htpb]
        \centering
		\begin{tabular}{|c|c|c|c|c|c|c|c|c|}
			\hline
			\diagbox{Strategy}{Time}{Dimension} & $2^8$ & $2^9$ & $2^{10}$ & $2^{11}$ & $2^{12}$ & $2^{13}$ & $2^{14}$ & $2^{15}$ \\
			\hline
			without MPI & 2s & 13s & 1m29s & 11m15s & 1h23m34s & -- & -- & -- \\
			\hline
			$2\times2$ & $<$1s & 6s & 40s & 4m45s & 35m46s & 4h37m53s & -- & -- \\
			\hline
			$4\times4$ & 11s & 43s & 2m46s & 11m14s & 49m43s & 3h54m38s & 20h13m58s & -- \\
			\hline
			$8\times8$ & 53s & 3m13s & 11m24s & 25m35s & 1h28m37s & 5h34m35s & 21h57m14s & 5d16h48m19s \\
			\hline
			$16\times16$ & 3m9s & 6m23s & 14m37s & 54m4s & 1h56m32s & 14h6m24s & 1d5m8s & 4d15h42m9s \\
			\hline
		\end{tabular}
		\caption{Comparison of time cost of distributed computing unitary evolution. The symbols "d", "h", "m" and "s" represent "day(s)", "hour(s)", "minute(s)" and "second(s)", respectively.}	
		\label{Tab:ComparisonTimeEvolution}
	\end{table*}
	
	\begin{figure*}
		\centering
		\includegraphics[width=1.\textwidth]{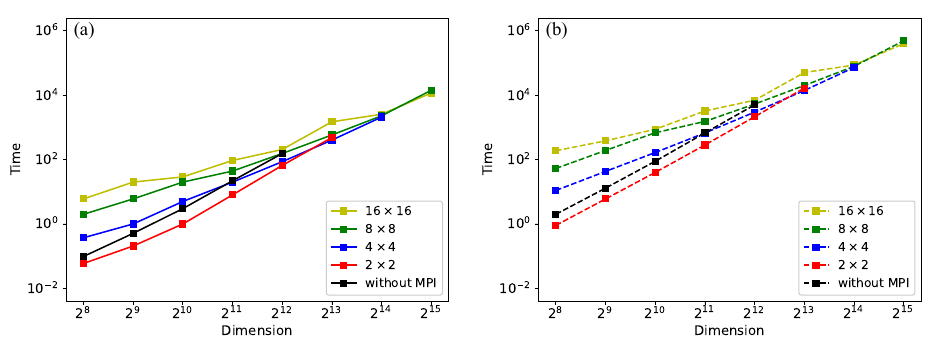} 
		\caption{(online color) Comparison of time cost. (a) shows the comparison of time cost of distributed computing Taylor series approximation, (b) shows the comparison of time cost of distributed computing unitary evolution.}
		\label{fig:ComparisonTime}
	\end{figure*}
	
	In this section, we study the interaction of photons with atoms group. We define the state of number of free photons as follows
	\begin{equation}
		\label{eq:PhotonState}
		\{m\succ=\sum_{\sum_{i=1}^n l_i=n-m}c_i\bigotimes_{i=1}^n|l_i\rangle
	\end{equation}
	where $m$ is number of free photons in the cavity, $c_i$ --- normalization factor. For example, $m=0$ means $\sum_{i=1}^n l_i=n-0=n$, that is to say, all atoms are in the excited state, that is, $l_1=l_2=\cdots=l_n=1$. If $m=n-1$, $\sum_{i=1}^n l_i=n-(n-1)=1$, that is to say, only one atom is in the excited state, and others are in the ground state.
	
	In Fig. \ref{fig:Evolution8}, we have eight atoms in the excited state, and the number of free photons in the cavity at the beginning is $0$. Thus, $n=8$ and the probability of $\{0\succ$, which is denoted by cyan solid curve, is equal to $1$. As time goes by, the curve of probability of $\{0\succ$ begins to drop from $1$ to $0$, and other curves begin to rise. The curve of probability of $\{1\succ$ reaches the peak first, then the curve of probability of $\{2\succ$ reaches the peak, and so on, and finally the curve of probability of $\{8\succ$ reaches the peak. In addition, these curves show obvious symmetry. The states are symmetric about $\frac{n}{2}=\frac{8}{2}=4$. For example, curves of probability of $\{0\succ$ and $\{8\succ$ have the same peak, curves of probability of $\{1\succ$ and $\{7\succ$ have the same peak, etc. In Fig. \ref{fig:EvolutionMore}, when $n=10,\ 12,\ 14$, we also get similar curves changes and the same symmetry.

	\subsection{Comparison of cost time and speedup of different distributed computing strategies}
	\label{subsec:Comparison}
	
	\begin{figure*}
		\centering
		\includegraphics[width=1.\textwidth]{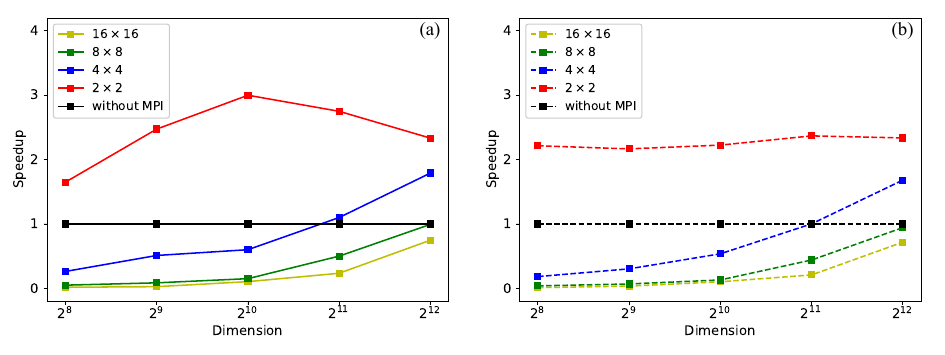} 
		\caption{(online color) Comparison of speedup. (a) shows the comparison of speedup of distributed computing Taylor series approximation, (b) shows the comparison of speedup of distributed computing unitary evolution.}
		\label{fig:ComparisonSpeedup}
	\end{figure*}
	
	We use the Cannon's algorithm to solve distributed computing quantum unitary evolution on the supercomputer platform. We have four strategies of distributed computing: the first strategy is to split each matrix into $2\times2=4$ blocks; the second strategy is to split each matrix into $4\times4=16$ blocks; the third strategy --- $8\times8=64$ blocks; the forth strategy --- $16\times16=256$ blocks. And we compare the time cost and speedup of tasks under these strategies with the situation without distributed computing.
	
	The comparison of time cost of distributed computing Taylor series approximation is shown in Tab. \ref{Tab:ComparisonTimeTaylor} and the comparison of time cost of distributed computing unitary evolution is shown in Tab. \ref{Tab:ComparisonTimeEvolution}. We found that when using $4$ computational cores for parallel computing, the time required to complete a task is always less than when no parallel computing strategy is used. However, the time cost of other strategies is more than that of the situation without distributed computing because of the high cost of data transmission between cores. However, as the dimension of the Hilbert space increases, the proportion of data transmission cost in the total time cost decreases, making the advantages of the strategy of high number of cores beginning to become obvious. Regarding the comparison of time cost, we can see it more intuitively in Fig. \ref{fig:ComparisonTime}.

	The comparison of speedup of distributed computing Taylor series approximation and distributed computing unitary evolution are shown in Tabs. \ref{Tab:ComparisonSpeedupTaylor} and \ref{Tab:ComparisonSpeedupEvolution}, respectively. For all dimensions, the average speedup under strategy with $4$ cores we can obtain is equal to $2$. For other strategies, as dimension increases, their speedup increases. When the dimension is equal to $2^{11}$, the speedup of the strategy with $16$ cores exceeds that of the situation without distributed computing. Similarly, when the dimension is equal to $2^{12}$, the speedup of the strategy with $64$ cores is almost close to that of the situation without distributed computing. Regarding the comparison of speedup, we can see it more intuitively in Fig. \ref{fig:ComparisonSpeedup}.
	
	\begin{table}[!htpb]
        \centering
		\begin{tabular}{|c|c|c|c|c|c|}
			\hline
			\diagbox{Strategy}{Speedup}{Dimension} & $2^8$ & $2^9$ & $2^{10}$ & $2^{11}$ & $2^{12}$ \\
			\hline
			without MPI & 1.000 & 1.000 & 1.000 & 1.000 & 1.000 \\
			\hline
			$2\times2$ & 1.650 & 2.474 & 3.000 & 2.750 & 2.333 \\
			\hline
			$4\times4$ & 0.262 & 0.510 & 0.600 & 1.100 & 1.791 \\
			\hline
			$8\times8$ & 0.049 & 0.085 & 0.150 & 0.500 & 0.994 \\
			\hline
			$16\times16$ & 0.016 & 0.026 & 0.103 & 0.234 & 0.748 \\
			\hline
		\end{tabular}
		\caption{Comparison of speedup of distributed computing Taylor series approximation.}	
		\label{Tab:ComparisonSpeedupTaylor}
	\end{table}	
	
	\begin{table}[!htpb]
        \centering
		\begin{tabular}{|c|c|c|c|c|c|}
			\hline
			\diagbox{Strategy}{Speedup}{Dimension} & $2^8$ & $2^9$ & $2^{10}$ & $2^{11}$ & $2^{12}$ \\
			\hline
			without MPI & 1.000 & 1.000 & 1.000 & 1.000 & 1.000 \\
			\hline
			$2\times2$ & 2.215 & 2.167 & 2.225 & 2.368 & 2.336 \\
			\hline
			$4\times4$ & 0.182 & 0.302 & 0.536 & 1.001 & 1.681 \\
			\hline
			$8\times8$ & 0.038 & 0.067 & 0.130 & 0.440 & 0.943 \\
			\hline
			$16\times16$ & 0.011 & 0.034 & 0.101 & 0.208 & 0.717 \\
			\hline
		\end{tabular}
		\caption{Comparison of speedup of distributed computing unitary evolution.}	
		\label{Tab:ComparisonSpeedupEvolution}
	\end{table}

	\section{Conclusion}
	\label{sec:Conclusion}
	
	We have implemented the distributed computing transformation of quantum unitary evolution on a supercomputing platform, and studied the TCM in high-dimensional Hilbert space. The results in Sec \ref{subsec:Interaction} show that the excitation-relaxation effects of atoms group in the TCM are periodic and symmetrical. In addition, we also analysed the impact of different distributed computing strategies (different numbers of computing cores) on the time cost of computing tasks and found the corresponding speedup. The results in Sec. \ref{subsec:Comparison} show that distributed computing with the help of the Cannon's algorithm can reduce memory usage and improve efficiency to a certain extent. In particular, the higher the dimension of the Hilbert space (the higher the dimension of the matrix), the smaller the proportion of data transmission cost to the total time cost becomes, and the more obvious the efficiency of distributed computing becomes.

	\begin{acknowledgments}
	The reported study was funded by China Scholarship Council, project number 202108090483. The authors acknowledge Center for Collective Usage of Ultra HPC Resources (https://www.parallel.ru/) at Lomonosov Moscow State University for providing supercomputer resources that have contributed to the research results reported within this paper.
	\end{acknowledgments}

	\bibliography{bibliography}

\end{document}